\documentclass[twocolumn,prl]{revtex4}
\usepackage{graphicx}
\usepackage{bm}
\usepackage{amsmath}
\begin{document} \title{Ferromagnetic imprinting of spin polarization in a
semiconductor}

\author{C. Ciuti, J. P. McGuire, and L. J. Sham}
\affiliation{Department of Physics, University of California San Diego, La
Jolla CA 92093-0319.}

\begin{abstract} We present a theory of the imprinting of the electron spin
coherence and population in an n-doped semiconductor which forms a
junction with a ferromagnet.  The reflection of non-equilibrium
semiconductor electrons at the interface provides a mechanism to
manipulate the spin polarization vector. In the case of unpolarized
excitation, this ballistic effect produces spontaneous electron spin
coherence and nuclear polarization in the semiconductor, as recently
observed by time-resolved Faraday rotation experiments.  We
investigate the dependence of the spin reflection on the Schottky
barrier height and the doping concentration in the semiconductor and
suggest control mechanisms for possible device applications.

\end{abstract}

\date{\today} \maketitle

The physics of the semiconductor/ferromagnet interface is a crucial
issue in the emerging field of spin electronics \cite{review}.
Progress has been made in injecting polarized electrons from
ferromagnets \cite{ohno,hammer,zhu,jonker3} or semimagnetic
semiconductors \cite{jonker,fied,ZnSe2,ghali} into semiconductors, an
essential step towards the exploitation of the spin degree of freedom
in a new generation of multifunctional devices.  Recently,
time-resolved optical Faraday rotation measurements have demonstrated
that the presence of a ferromagnetic interface
\cite{FMimprinting,FMcoherence} can produce spontaneous electron spin
coherence and large nuclear magnetic fields in the
semiconductor. Moreover, the spin-polarized photocurrent from a
semiconductor into a ferromagnet \cite{Cambridge,isakovic} has been
shown to vary with optical polarization. We report here a theoretical
study of the spin reflection of non-equilibrium electrons at the
semiconductor/ferromagnet interface. Our results show how (1)
unpolarized electrons {\em in the semiconductor} are spontaneously
polarized and (2) a pre-existing spin polarization vector is
tilted. The underlying physics is reminiscent of the Mott scattering
\cite{mott} of electrons with atoms via spin-dependent interaction.

Our theory is based on the general scattering theory of the electron
spin density matrix \cite{mott}. In particular, we focus on tunneling
through an insulator layer \cite{slon,jansen} or a Schottky barrier
\cite{rashba}.  The source of the electrons may be electrical or
optical.  The non-equilibrium carriers produce a spin current into the
ferromagnet during the short momentum relaxation time following a
pulsed excitation. This ultrafast ballistic process leaves a net spin
coherence and population in the semiconductor.  We show how to extract
from the time-resolved Faraday rotation the ballistic spin transport
properties of the semiconductor/ ferromagnetic junction and predict
the dependence of the imprinted spin on the system parameters.

\begin{figure}[t!]
\includegraphics[scale=0.41]{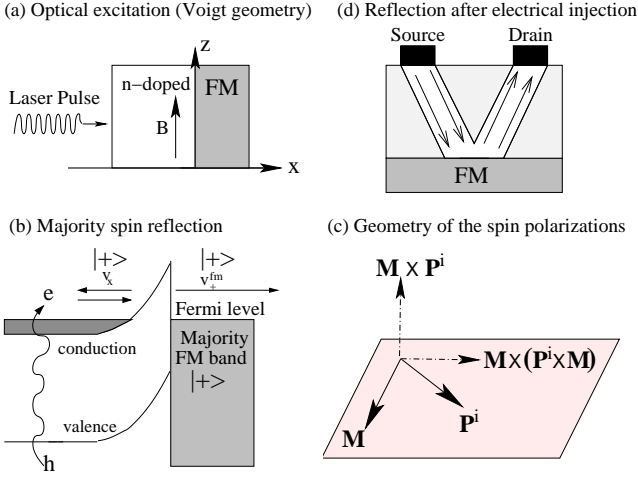}
\caption{Counterclockwise: (a) Sketch of the semiconductor/ferromagnet 
junction excited by a light beam in Voigt configuration. 
(b) Schematic band diagram for
the reflection and transmission of a semiconductor electron in the majority
spin channel ($|+\rangle$ is the spin state of the ferromagnetic majority band). For
the minority spin band the diagram is analogous, but with a
different Fermi velocity. 
(c) Geometry involved in the ferromagnetic imprinting
when the pumped electrons have a spin polarization ${\bf P}^{\rm i}$. The
polarization of the reflected electrons will have also a component along
$\hat{\bf M} \times {\bf P}^{\rm i}$, i.e. orthogonal to the ferromagnet
magnetization and the incident polarization (see Eq. (\ref{polar})).  When
${\bf P}^{\rm i} = 0$, the imprinted spin is directed along $\hat{\bf M}$.
(d) Sketch of electrical device where ballistic electrons are injected
though lateral contacts.}
\label{sketch}
\end{figure}

The Hamiltonian for the semiconductor/ferromagnet junction (see Fig.
\ref{sketch}a-b),
\begin{equation}
\mbox{\hspace{-0.1in}}  H = K + V(x) +
\frac{g^{\star}}{2}  \mu_{\rm B} {\bm \sigma} \cdot {\bf B}_{\rm T}
\Theta (-x) +\frac{\Delta}{2} \sigma_M \Theta (x),
\end{equation} consists of: the kinetic energy $K$ for the conduction band on
the semiconductor side and two spin bands with exchange splitting
$\Delta$ in the ferromagnet; the Schottky barrier potential $V(x)$;
the Zeeman energy in the semiconductor where $g^{\star}$ is the
effective electron g-factor ($g^{\star}_{\rm GaAs} = - 0.44$),
$\mu_{\rm B}$ the Bohr magneton, and ${\bm \sigma}$ the vector of
Pauli matrices.  The total field ${\bf B}_{\rm T}$ is the sum of the
applied field ${\bf B}$ and of the local field ${\bf B}_{\rm N}$ due
to the nuclear polarization.  $\sigma_M$ is the Pauli matrix in the
magnetization direction and $\Theta$ is the step function. The orbital
effect of the weak magnetic field is neglected.

The contribution to the spin polarization by the valence holes can be
neglected since the hole spin relaxation time is much shorter than the
optical recombination time $T_{\rm rec}$ \cite{holes} due to spin
mixing in the valence bands by the spin-orbit
interaction\cite{ueno}. The spin dephasing time of the conduction
electron in n-doped samples is shown to be much longer than the
recombination time \cite{longlived}. Because of the dominant role of
the magnetization in the ferromagnet, we denote the majority and
minority spin states by $|+ \rangle$ and $|-\rangle$
respectively. These states are eigenstates of the ferromagnetic
exchange splitting operator, namely $\sigma_M|\pm\rangle =
\mp|\pm\rangle$ (note that the magnetization ${\bf M}$ is antiparallel
to the net electron spin ${\bf S}^{\rm fm}$). The spin-dependent
reflection of the semiconductor electrons at the ferromagnetic
interface is represented by the reflection matrix $\hat{r}({\bf k})$
in the electron spin space. In the ferromagnetic spin basis $\{ |-
\rangle,|+ \rangle \}$, the reflection matrix has the diagonal
representation $ \hat{r}({\bf k}) = |-\rangle r_{-,{\bf k}} \langle -
|+ |+ \rangle r_{+,{\bf k}} \langle + |$, where $r_{-,{\bf k}}$
($r_{+,{\bf k}}$) is the reflection coefficient for a semiconductor
electron spin aligned with the ferromagnet minority (majority) spin
band. This diagonal representation in the ferromagnetic spin basis is
exact only if the semiconductor magnetic field ${\bf B}_{\rm T}$ is
parallel to the magnetization ${\bf M}$. However, the exchange
splitting $\Delta$ in a ferromagnet is typically several orders of
magnitude larger than the semiconductor Zeeman splitting. We have
verified explicitly that the effect of the magnetic field ${\bf
B}_{\rm T}$ on the spin-dependent reflection matrix is negligible.
Also, the effective magnetic field due to the lack of the inversion
symmetry (the Rashba effect) is negligible here.  For extraction of
the spin polarization, it is convenient to express the reflection
matrix in the form
\begin{equation}
\label{general}
\hat{r}({\bf k}) = \frac{1}{2}\left[ (r_{-,{\bf k}} + r_{+,{\bf k}})
\openone +  (r_{-,{\bf k}} - r_{+,{\bf k}})  \hat{\bf M} \cdot {\bm
\sigma} \right] ,
\end{equation} where $\openone$ is the unit matrix and $\hat{\bf M}$ is the
unit vector along the magnetization.

Consider the excitation of the n-doped semiconductor by a short
optical pulse.  Immediately after the light absorption, the 
excited electron
spin density matrix on the semiconductor side is $\hat{\rho}^{\rm
i}({\bf k},t=0) =f^{\rm i}(k) \frac{1}{2} (\openone + {\bf P}^{\rm i}
\cdot {\bm \sigma} )$, where $f^{\rm i}(k)$ is the
non-equilibrium electron distribution injected by the exciting
laser. The polarization vector ${\bf P^{\rm i}}$ depends on the optical
selection rules. Since the pump has injected electrons, 
there will be an electron current (per unit area) from the semiconductor
into the ferromagnet,
\begin{equation}
\hat{j}(t) = \hat{j}^{\rm i}(t) + \hat{j}^{\rm r}(t) = \int_{k_{\rm x} > 0}
\frac{d^{3} {\bf k}}{(2 \pi)^3} [\hat{\rho}^{\rm i}({\bf k},t) -
\hat{\rho}^{\rm r}({\bf k},t)] v_x ,
\end{equation} where $v_x =\hbar k_x/m_s$ is the  velocity component 
normal to the interface. The reflected density matrix is
given by
\begin{eqnarray}
\hat{\rho}^{\rm r}({\bf k},t) &=&  \hat{r}({\bf k}) \hat{\rho}^{\rm i}({\bf
k},t) \hat{r}^{\dagger}({\bf k}) \\ &=& f^{\rm i}(k,t) \frac{1}{2} [R_0({\bf k})
\openone + {\bf R}({\bf k}) \cdot {\bm \sigma} ],
\end{eqnarray} where, with the ${\bf k}$-dependence understood,
\begin{eqnarray} R_0 &=& \frac{1}{2}[(|r_-|^2 + |r_+|^2) + (|r_-|^2 - |r_+|^2)
~\hat{\bf M} \cdot {\bf P}^{\rm i}] , \label{r0} \\ {\bf R} &=& \frac{1}{2}[ (|r_-|^2
- |r_+|^2) + (|r_-|^2 + |r_+|^2)~
\hat{\bf M} \cdot {\bf P}^i ] ~\hat{\bf M} \nonumber \\ &+& {\rm Re}(r_-
r_+^*)~ (\hat{\bf M}\times {\bf P}^{\rm i} )\times \hat{\bf M} - {\rm Im}(r_-
r_+^*)~\hat{\bf M} \times {\bf P}^{\rm i}. \nonumber
\\ \label{polar}
\end{eqnarray}
The time-dependence of $\hat{\rho}^{\rm i}({\bf k},t)$ is dominated by
the relaxation of the hot carrier distribution $f^{\rm i}(k,t)$.  The
spin precession due to a weak applied magnetic field and spin
relaxation occur on a much longer time scale compared to the fast
orbital relaxation. In the relaxation time approximation, the decay of
the non-equilibrium electron population is $f^{\rm i}(k,t) = f^{\rm i}(k)
\exp(-t/\tau_k)$, yielding a simple expression for the surface
spin density imprinted in the semiconductor,
\begin{eqnarray} {\bf S^{\rm r}} &=& - {\rm Tr} \left\{\frac{\hbar}{2}{\bm \sigma}
\int dt~ [\hat{j}^{\rm i}(t) + \hat{j}^{\rm r}(t)] \right\} \nonumber \\
&= & \frac{\hbar}{2} ~
\int_{k_{\rm x} > 0}
\frac{d^{3} {\bf k}}{(2 \pi)^3} f^{\rm i}(k) \left [ {\bf R}({\bf k}) - 
{\bf P}^{\rm i}\right ] \tau_k v_x~. \end{eqnarray} 
After the transient, the surface spin density inherited by the n-doped semiconductor is
${\bf S^{\rm tot}} =  n^{i}~L~\frac{\hbar}{2}~{\bf P}^{\rm i} + {\bf S^{\rm r}}$ where
$n^{\rm i} = \int \frac{d^{3}{\bf k}}{(2 \pi)^3} f^{\rm i}(k)$ is the volume
density of pumped electrons and $L$ is the sample length. 
Of course, this holds when $n^{\rm i}$ is smaller than the doping
density $n$ (acting as a spin reservoir)
 and when $L$ is larger than the mean-free path. 
In fact, for very thin samples, our treatment needs to be refined to
account for multiple reflections.

In the following, we will consider separately the
two cases of spin unpolarized and polarized excitation.
\\ {\it (i) Unpolarized excitation} \\ The injected electron ensemble
is unpolarized (${\rm P}^{\rm i} = 0$). Reflection from the
ferromagnet results in a net spin surface density,
\begin{equation} {\bf S}^{\rm r} =    \frac{\hbar}{4} ~\hat{\bf M}
\int_{k_{\rm x} > 0}
\frac{d^{3} {\bf k}}{(2 \pi)^3} f^{\rm i}(k) \left(|r_{-,{\bf k}}|^2-
|r_{+,{\bf k}}|^2 \right) \tau_k v_x~. \label{result}
\end{equation} The amplitude of the imprinted spin in the semiconductor
is determined by the difference between the spin-dependent
reflectivities.  For each momentum channel the mean free path is
$\tau_k v_x$.  Finally, ${\bf S}^{\rm r}$ is aligned either parallel
or antiparallel to the ferromagnetic magnetization ${\bf M}$,
depending on the interface properties, which will be shown below.

In an applied magnetic field ${\bf B}$, the Larmor precession of the
imprinted spin polarization is given by $\partial_t {\bf S}(t) =
\frac{1}{\hbar} g^{\star} \mu_{\rm B} {\bf B}_{\rm T} \times {\bf
S}(t) -\frac{1}{T_2}~ {\bf S}(t) $, where $T_2$ is the spin relaxation
time and the initial condition is ${\bf S}(t \approx 0)={\bf S}^{\rm
r}$.  The total field ${\bf B}_{\rm T} = {\bf B} + {\bf B}_{\rm N}$
contains the local contribution ${\bf B}_{\rm N}$ due to the nuclear
polarization \cite{orientation} where ${\bf B}_{\rm N} \sim
(g^{\star}/|g^{\star}|) ({\bf S}^r \cdot {\bf B})~{\bf B}/B^2$. The
imprinting of the nuclear spins also affects the effective Larmor
precession frequency $\Omega_{L} = g^{\star} \mu_{\rm B} (B+B_{\rm N})
/ \hbar$ \cite{NMR} which can be measured by the time-resolved optical
Faraday rotation \cite{FMimprinting,FMcoherence}.  The Faraday
rotation of a linearly polarized probe beam is proportional to the
component of the net electron spin along the direction of the photon
wave-vector ${\bf k}_{\rm phot}$ in the medium, i.e.  $ {\rm FR}(t)
\sim {\bf S}(t) \cdot {\bf k}_{\rm phot}/k_{\rm phot} $.  Now,
consider the Voigt geometry with light propagating along the x
direction and magnetic field in the $z$ direction. Let the
ferromagnetic magnetization be in the interface plane at an angle
$\alpha$ to ${\bf B}$ (i.e. ${\bf M} = M (0,\sin \alpha, \cos
\alpha)$).  Since $g^{\star} < 0$, the nuclear field is $ B_{\rm N}
\sim \langle \left ( |r_{+}|^2 - |r_{-}|^2 \right ) \tau v_x \rangle
\cos \alpha $.  The spontaneous Faraday rotation is $ {\rm FR}(t) \sim
\langle \left ( |r_{-}|^2 - |r_{+}|^2 \right ) \tau v_x \rangle \sin
\alpha \sin (\Omega_L t ) e^{-t/T_2}.  $ Therefore, for $\alpha = 0$
the nuclear imprinting is maximum while there is no spontaneous
Faraday rotation. On the other hand, for $\alpha = \pi /2$, the
nuclear field vanishes while the spontaneous coherence has the maximum
amplitude. Moreover, the Faraday rotation oscillates in time as $\sin
(\Omega_L t)$. These trends are in perfect agreement with the
experimental observations \cite{FMimprinting,FMcoherence}.  The theory
predicts similar spin polarization effects when an unpolarized current
is injected electrically, for example through a ballistic V-groove
with non-magnetic contacts as shown in Fig.~\ref{sketch}d.

\noindent {\it (ii) Polarized excitation} \\ When the excited electron
population has a pre-existing spin polarization (${\bf P}^{\rm i}\neq
{\bf 0}$), the current in the semiconductor is changed by the spin
polarization and the spin polarization is altered. In addition to two
components in the plane defined by the magnetization and initial
polarization vectors, there will be an additional component normal to
the plane [cf. Eq.~(\ref{polar})], as shown in Fig.~1c. Evidence for
polarization-dependence of the ferromagnetic imprinting has been
reported in Ref.~\onlinecite{FMcoherence}. Additional studies as a
function of the relative orientation between $\hat{\bf M}$ and ${\bf
P}^{i}$ will be needed to test the dependence predicted in
Eq.~(\ref{polar}).  Such a change of the spin polarization direction
could provide a mechanism for the manipulation of the semiconductor
spin polarization vector.

For the phenomena discussed above to be measurable, the
spin-dependence of the reflection coefficients need to be sufficiently
strong. We point out that since the transmittance $|t|^2 = 1 - |r|^2$
in either direction through the barrier is the same, reflection in the
semiconductor is no more efficient in creating spin polarization than
injection from the ferromagnet.  To investigate the dependence on the
parameters of the Schottky barrier, we calculate the spin-dependent
reflection coefficients for a simplified model of the
semiconductor/ferromagnet junction.  We consider first a rectangular
barrier $V(x < 0) = U_{\rm b} \Theta (x+a) -E_{\rm s} \Theta (-x-a)$,
where $U_{\rm b}$ is the barrier height measured from the Fermi level
and $E_{\rm s} = \hbar^2 (3 \pi^2 n)^{2/3}/(2 m_{\rm s})$ is the
kinetic Fermi energy of the n-doped semiconductor. The spin-dependent
reflection coefficients are
\begin{eqnarray} r_{\pm,\bf k} =
\frac{e^{2k_{\rm b}a} (iv^{\rm fm}_{\pm}-v_{\rm b})(iv_{\rm x} + v_{\rm b}) -
(iv^{\rm fm}_{\pm}+v_{\rm b})(iv_{\rm x} - v_{\rm b})} {e^{2k_{\rm b}a}
(iv^{\rm fm}_{\pm}-v_{\rm b})(iv_{\rm x} - v_{\rm b}) - (iv^{\rm
fm}_{\pm}+v_{\rm b})(iv_{\rm x} + v_{\rm b})}~, \nonumber
\nonumber
\end{eqnarray} where $v^{\rm fm}_{\pm}$ is the Fermi velocity (along the
x-direction) of  the majority (minority) band in the ferromagnet. In the
simplified case of two parabolic bands, $v^{\rm fm}_{+} = \sqrt{2 E_{F}/m_{\rm
fm}}$  and $v^{\rm fm}_{-}=
\sqrt{2 (E_{F}-\Delta)/m_{\rm fm}}$, implying that $v_{+}^{\rm fm}  >
v_{-}^{\rm fm}$. Note that a non-parabolic band dispersion can lead  to an
opposite relationship, i.e. $v_{+}^{\rm fm}  < v_{-}^{\rm fm}$. The wave-vector
$k_{\rm b} = \sqrt{ 2m_{\rm s}~(U_{\rm b}+E_{\rm s})/\hbar^2 - k_{\rm x}^2}$ is that of the
evanescent wave in the barrier region and
$v_{\rm b} = \hbar k_{\rm b}/m_{\rm s}$.  Finally, $v_{\rm x} = \hbar k_{\rm
x}/m_{\rm s}$ is the incident velocity in the semiconductor. A more realistic
Schottky barrier is represented by a parabolic bent potential $V(x)$
(see Fig. 1b) with depletion layer
$d \approx
\sqrt{\epsilon_0 U_{\rm b}/(2 \pi n e^2)}$ where $\epsilon_0$ the dielectric
constant.  To improve our estimate, for each value of $k_{x}$ we
approximate the reflection coefficients by taking an effective rectangular
barrier width $a= a(k_{\rm x})$ such that the action integral in the barrier
region is the same.

\begin{figure}[t!]
\includegraphics[scale=0.575]{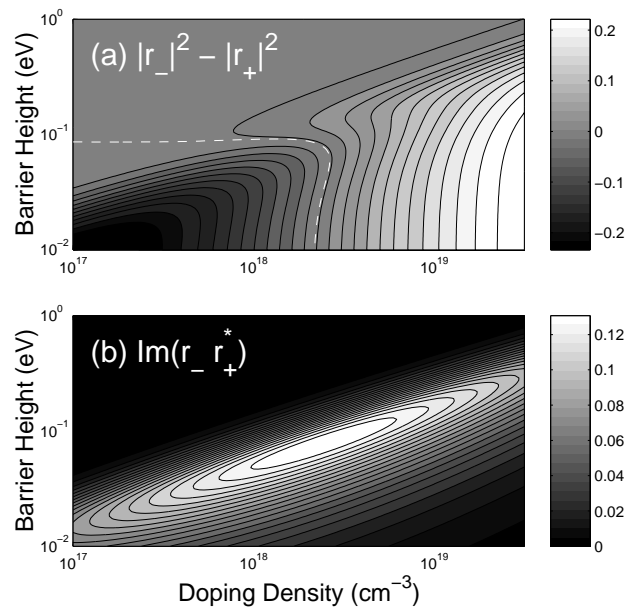}
\caption{(a) Contours of $|r_{-,{\bf k}}|^2 - |r_{+,{\bf k}}|^2$ as a
function of the doping density $n$ (log scale) and the Schottky
barrier height $U_{\rm b}$ (log scale).  For each density, ${\bf k}$
is taken along the x-direction with amplitude $(3 \pi^2 n)^{1/3}$, the
Fermi wavevector in the n-semiconductor.  The semiconductor parameters
are those of bulk GaAs.  Ferromagnet Fermi velocities: $v^{\rm fm}_+ =
9.4 \times 10^{7} {\rm cm}/{\rm s}$, $v^{\rm fm}_- = 4.6 \times 10^{7}
{\rm cm}/{\rm s}$ (corresponding to two exchange split parabolic bands
with $E_{F} = 2.5$~eV, $\Delta = 1.9$~eV and free-electron mass).  At
the white-dashed line, $|r_{-,{\bf k}}|^2 - |r_{+,{\bf k}}|^2$ changes
sign.  (b) Contours of ${\rm Im}{(r_- r_+^{\star} })$.  ${\rm Im}{(r_-
r_+^{\star} })$ has always the same sign.}
\label{diagram}
\end{figure} 
Representative results for realistic material parameters are plotted
in Fig.~\ref{diagram}. Panel (a) shows contours of $|r_{-,\bf
k}|^2-|r_{+,\bf k}|^2$ as a function of the doping density $n$ and the
barrier height $U_b$.  In the region where tunneling is significant,
the predicted difference between majority and minority reflectivities
can be 25 \% (the polarization is actually large also when tunneling
is small \cite{remark}).  When $|r_{-,\bf k}|^2-|r_{+,\bf k}|^2$ is
positive (negative), the imprinted spin ${\bf S}^{\rm r}$ is parallel
(antiparallel) to the magnetization ${\bf M}$ in the ferromagnet. For
vanishing barrier height and low doping, the transmittance into the
ferromagnet is dominated by the spin channel with the lower
ferromagnet Fermi velocity, allowing a better matching with the small
semiconductor velocity. However, with increasing doping, the
semiconductor velocity increases and eventually can match better the
ferromagnet spin band with the larger velocity.  This is the reason
for the change of sign of $|r_{-,\bf k}|^2-|r_{+,\bf k}|^2$ as a
function of doping in the low barrier region. When the barrier height
is large enough, the velocity matching is not the only relevant
feature.  Indeed, for moderate doping a barrier increase can induce a
change of sign of $|r_{-,\bf k}|^2-|r_{+,\bf k}|^2$ in analogous way
to what is predicted for ferromagnet/oxide junctions \cite{slon}. This
prediction suggests that by using a forward bias (i.e., semiconductor
Fermi level higher than that in the metal) to tune the semiconductor
barrier $U_{\rm b}$, the sign and amplitude of the imprinted spin
could be controlled.

  Actually, different spin orientations with different materials have
been observed experimentally \cite{FMcoherence}.  In a GaAs/MnAs
heterojunction, ${\bf S}^{\rm r}$ has been found to be antiparallel to
${\bf M}$, while for a GaAs/Fe system the spin is parallel. Recent
transport characterizations of GaAs/MnAs suggest that the Schottky
barrier in this system can be very low, allowing nearly ohmic
conduction \cite{MnAsohmic} and significant spin valve effect in
trilayer structures \cite{MnAsvalve}.  In the case of iron, the
barrier is quite high (0.7 eV) and thus our model predicts an
imprinted spin parallel to ${\bf M}$ as observed \cite{FMcoherence}.
The relatively high efficiency observed in the case of the iron
junction at very low doping density ($\approx 10^{17} {\rm cm}^{-3}$)
\cite{FMcoherence} may be due to the very thin GaAs active region ($L
= 100 ~{\rm nm}$) of the investigated sample. The effects of finite
size on the Schottky barrier and of multiple electron reflections
could significantly enhance the spin reflection efficiency, an issue
under current investigation.  To complete our study, Fig.  2b depicts
the contours of ${\rm Im}(r_- r^{\star}_+)$, representing the
amplitude of the component of the imprinted spin along $\hat{\bf M}
\times {\bf P}^i$ (see Eq. (\ref{polar})) which we predict to occur
when the excited electrons are already polarized (${\bf P}^i \neq {\bf
0}$).

In conclusion, we have presented a ballistic theory for the
ferromagnetic imprinting of the spin coherence and population in an
n-doped semiconductor following pulsed excitation.  We have given an
analysis of the optical pumping and the all-optical detection of spin
transport through the time-resolved Faraday rotation, giving a
framework for the interpretation of recent experiments
\cite{FMimprinting,FMcoherence}. We suggest that an electrical tuning
of the Schottky barrier could allow for control of the sign and
amplitude of the imprinted spin.  For the possibilities of device
applications, generation of spin currents and reorientation of the
spin polarization are predicted also for electrical excitation of
non-equilibrium electrons in the semiconductor using traditional
non-magnetic contacts. Detailed studies of particular systems and
quantitative comparison with experiments will be given elsewhere.

\begin{acknowledgments} This work is supported by DARPA/ONR N0014-99-1-1096 and
NSF DMR 0099572. CC is also grateful to the Swiss National Foundation for
additional support. JPM acknowledges a graduate fellowship
by California Institute for Telecommunications and Information
Technology.  We thank D.D. Awschalom for 
discussions and for the experimental results \cite{FMcoherence} prior
to publication.

\end{acknowledgments}


\begin{thebibliography}{}

\bibitem{review} S.A. Wolf {\it et al.}, Science {\bf 294},
1488 (2001).
\bibitem{ohno} Y. Ohno, {\it et al.}, Nature {\bf 402}, 790 (1999).
\bibitem{hammer} P.R. Hammer, B.R. Bennett, M.J. Yang, and M. Johnson, Phys.
Rev. Lett. {\bf 83}, 203 (1999) and {\bf 84}, 5024 (2000). F.G. Monzon, H.X.
Tang, and M.L. Roukes, {\em ibid} {\bf 84}, 5022 (2000). B.J. van Wees, {\em
ibid} {\bf 84}, 5023 (2000).
\bibitem{zhu} H.J. Zhu, {\it et al.}, Phys. Rev. Lett. {\bf 87}, 016601 (2001).
\bibitem{jonker3} A.T. Hanbicki, {\it et al.}, Appl. Phys. Lett. {\bf 80}, 1240
(2002).
\bibitem{jonker} B.T. Jonker, U.S. Patent No. 5,874,749 (filed in 1993, awarded
in 1999). B.T. Jonker, {\it et al.}, Phys. Rev. B {\bf 62}, 8180 (2000).
\bibitem{fied} R. Fiederling, {\it et al.}, Nature {\bf 402}, 787 (1999).
\bibitem{ZnSe2} I. Malajovich, {\it et al.}, Nature {\bf 411}, 770 (2001).
\bibitem{ghali} M. Ghali, {\it et al.}, Solid State Commun. {\bf 119}, 371
(2001).
\bibitem{FMimprinting}  R.K. Kawakami {\it et al.}, Science {\bf 294}, 131
(2001).
\bibitem{FMcoherence} R.J. Epstein {\it et al.}, to be published in Phys. Rev.
B; preprint cond-mat/0201350.

\bibitem{Cambridge} A. Hironata {\it et al.}, Phys. Rev. B {\bf 63}, 104425
(2001).
\bibitem{isakovic} A.F. Isakovic, {\it et al.}, Phys. Rev. B {\bf 64}, 161304
(2001).
\bibitem{mott} N.F. Mott and H.S.W. Massey, {\em The theory of atomic
collisions}, third edition (Oxford University Press, London, 1965), Chapter X.
\bibitem{slon} J.C. Slonczewski, Phys. Rev. B {\bf 39}, 6995 (1989).
\bibitem{jansen} R. Jansen, M.W.J. Prins, and H. Van Kempen, Phys. Rev. B {\bf
57}, 4033 (1998).
\bibitem{rashba} E.I. Rashba, Phys. Rev. B {\bf 62}, R16267 (2000).
\bibitem{holes} T.C. Damen {\it et al.}, Phys. Rev. Lett. {\bf 67}, 3432
(1991).
\bibitem{ueno} T. Uenoyama and L.J. Sham, Phys. Rev. Lett. {\bf 64}, 3070
(1990).
\bibitem{longlived} J.M. Kikkawa and D.D. Awschalom, Phys. Rev. Lett. {\bf 80},
4313 (1998).
\bibitem{orientation} F. Meier, B.P. Zakharchenya, Eds., {\it Optical
Orientation} (North-Holland, New York, 1984) and references therein.
\bibitem{NMR} G. Salis {\it et al.}, Phys. Rev. Lett. {\bf 86}, 2677 (2001).
\bibitem{remark} The traditional tunneling current polarization is $(|t_{+}|^2-|t_{-}|^2)/
(|t_{+}|^2+|t_{-}|^2) = (|r_{-}|^2-|r_{+}|^2)/
(2-|r_{-}|^2-|r_{+}|^2)$. In the high barrier regime, this
quantity can be as high as 30 \% (not shown) because $|r_{-}|^2 \approx |r_{+}|^2 \approx
1$.
\bibitem{MnAsohmic} W. Van Roy {\it et al},
 J. Cryst. Growth {\bf 227}, 852 (2001).
\bibitem{MnAsvalve} M. Tanaka, K. Takahashi, J. Cryst. Growth {\bf 227}, 847
(2001).

\end{thebibliography}
\end{document}